\documentclass[twocolumn]{revtex4-1}

\usepackage{graphicx}
\usepackage{dcolumn}
\usepackage{bm}

\usepackage[dvipsnames]{xcolor,colortbl}
\usepackage{tabularx}
\usepackage{array}
\usepackage{amsmath}

\newcolumntype{P}[1]{>{\centering\arraybackslash}p{#1}}
\definecolor{Gray}{gray}{0.85}
\definecolor{LightCyan}{rgb}{0.88,1,1}

\begin{document}

\preprint{APS/123-QED}

\title{
Static and dynamic properties of a binary, symmetric mixture of ultrasoft particles in the vicinity of criticality}

\author{Tanmay Biswas}
\email{tanmay.biswas@tuwien.ac.at}
\affiliation{Institut f\"ur Theoretische Physik, TU Wien, Wiedner Hauptstra{\ss}e 8-10, A-1040 Wien, Austria}

\author{Gerhard Kahl}%
 \email{gerhard.kahl@tuwien.ac.at}
\affiliation{Institut f\"ur Theoretische Physik, TU Wien, Wiedner Hauptstra{\ss}e 8-10, A-1040 Wien, Austria}

\author{Gaurav P. Shrivastav}
\email{gaurav.shrivastav@tuwien.ac.at}
\affiliation{Institut f\"ur Theoretische Physik, TU Wien, Wiedner Hauptstra{\ss}e 8-10, A-1040 Wien, Austria}

\date{\today}

\begin{abstract}
We investigate the static and the dynamic properties of an binary, equimolar, size-symmetric mixture of ultrasoft particles in the vicinity of the critical point of the system. Based on the generalized exponential potential (GEM) of order four for the particle interaction and using extensive molecular dynamics simulations in the canonical ensemble we investigate the above mentioned properties for various scenarios: we consider several  super- and subcritical states, we expose the system to rapid quenches and to external shearing forces. Based on an accurate determination of the phase diagram and of the location of the critical point we study the static structure of the system in terms of particle-based radial distribution functions. As systems of GEM particles are prone to cluster formation we complement these investigations by a detailed analysis of the composition of the clusters and of their spatial correlations for the different scenarios introduced above.  Furthermore we analyse the temperature dependence of the diffusivity of the particles and of the shear viscosity of the system. All these data provide a detailed and profound insight into the properties of the system under phase separation conditions and near criticality.
\end{abstract}

\maketitle


\section{Introduction}
\label{sec:introduction}

The phenomenon of liquid-liquid phase separation is ubiquitous in mixtures of soft matter systems \cite{hyman2014liquidphaseseparation, bates1991polymerphase} and often represents an important route to stir self-assembly processes in such system. Typically such a phase separation is induced by thermal quenches of equilibrated, supercritical mixtures down to subcritical temperatures. While numerous related investigations have been dedicated during the past years to archetypical liquid systems (such as Lennard-Jones -- LJ -- mixtures \cite{subir2006static, subir2003transport, shaista2007crossover, shaista2010kinetics}, only a few studies have dealt with typical (ultra-)soft particles, i.e., a class of molecular entities that can (partly or fully) overlap at an energy cost which is comparable to thermal energies.

In an effort to fill the above mentioned gap we recently started \cite{Biswas2024} with the help of molecular dynamics (MD) simulations our investigations on the phase behaviour and the separation kinetics of binary mixture of ultrasoft particles, which interact via the so called ``generalized exponential model of index $n$'' (GEM$n$), a meanwhile widely used model potential for ultrasoft particles (which fulfills the above mentioned features of particle overlap) 
\cite{ mladek2006formation, mladek2007phase, mladek2008multiple, likos2007ultrasoft, nikoubashman2011cluster, coslovich2011hopping, zhang2010reentrant}. In an effort to restrict the dimensionality of the space of the system parameters at a reasonable size and following existing studies on binary LJ systems \cite{subir2006static, subir2003transport, shaista2007crossover, shaista2010kinetics} we have considered a binary, equimolar, size-symmetric of GEM particles: the interactions between like particles are identical while the cross-interaction (which is stronger than the like-interaction) is proportional to the like interaction, quantified by the parameter $\zeta$ ($> 1$). In this contribution we track how different types of properties (static structure, diffusivity, transport coefficients) change as we cool the system from a homogeneous, supercritical phase down to temperatures which lie (slightly) below the critical temperature. 

In this sense the present contribution represents a natural extension to a previous publication  \cite{Biswas2024}: based on  extensive moleculardynamics (MD) simulations (predominantly in the canonical ensemble) we focus here on the properties of the system in the near-critical region. The first challenge we have to face is to find the exact location of the critical point and of the coexistence lines, determined via analysing in the vicinity of the critical temperature the probability distribution of the local concentration of the two species, since these properties depend in a sensitive manner on the size of the simulation cell: a detailed quantitative analysis of the phase diagram in temperature vs. density space of the system reveals that only for relatively large ensembles (i.e., for at least $\sim$ 65~000 particles) ensembles of different size yield consistent results of the location of the critical point. For ensemble of (at least) this size we have first calculated the structure in terms of (partial) radial distribution functions and number-number -- $S_{\rm nn}(q)$ -- and of concentration-concentration -- $S_{\rm cc}(q)$ -- structure factors, the latter one being analysed in terms of its well-known diverging behaviour close to criticality; to be more specific we have made an analysis based on the the quantity $\lim_{q \to 0} [1/S_{\rm cc}(q)]$ within a so-called  Ornstein-Zernike plot to determine the location of the critical point which reproduces within high numerical accuracy the critical point, determined via the concentration distributions. Further the structure is analysed in terms of particle- and cluster-based radial distribution functions (RDFs). With this information at hand we proceed to the transport coefficients, notably the diffusion constant $D(T)$, calculated both via the mean square displacement and the velocity autocorrelation function. As detailed in the manuscript $D(T)$ follows an Arrhenius type behaviour; we have calculated the characteristic activation energy of the GEM mixture which has then been compared to the related value of a size-symmetric, equimolar Lennard-Jones mixture. Via the Green-Kubo formalism we have calculated the shear viscosity which -- together with the diffusion constant -- leads to the Stokes-Einstein diameter; this quantity is considerably slower as the respective value in a related LJ mixture. Eventually we have analysed the the temperature dependence of the viscosity near criticality which shows in this region the predicted power law.

The manuscript is organized as follows: in the subsequent Section we briefly present the model and the methods, referring the interested reader for methodological details to the previous contribution \cite{Biswas2024}. In Section III we discuss the results: we start with the phase diagram, we then focus on the structural properties of the system (in terms of radial distribution functions and structure factors), we proceed to the diffusion properties and the viscosity and eventually discuss the properties of the system under shear. The manuscript is closed with concluding remarks and an outlook to ongoing and future work.

\section{Model and methods}
\label{sec:model_methods}

The investigations of this contribution are based on exactly the same model and the same simulation methods as in a preceding study on the phase separation dynamics of a binary mixture of ultrasoft particles \cite{Biswas2024}. Thus we will keep this section rather short, referring the interested reader to the previous publication.

We consider a binary, equimolar, size-symmetric mixture of ultrasoft particles (with species A and B) which interact via the generalized exponential model of index $4$ (GEM4), given by \cite{mladek2006formation, Biswas2024, gaurav2020, gaurav2021softmatter, caprini2018cluster, coslovich2012clusterglass}: 

\begin{equation}
    \Phi_{\alpha \beta}(r) = \epsilon_{\alpha \beta} \exp[-(r/\sigma_{\alpha \beta})^4] ,
    \label{gem4-equation} 
\end{equation}
with -- henceforward -- $\alpha, \beta$ = A or B. Due to the size-symmetry there is only one single length scale $\sigma  \equiv \sigma_{\alpha \beta}$; further, equimolarity imposes the relation $x_{\rm A} = x_{\rm B} = 1/2$ for the concentrations of each species. $N = N_{\rm A} + N_{\rm B}$ (with $N_{\rm A} = N_{\rm B}$) is the total number of particles in the ensemble. Further we choose the energy-parameters $\epsilon_{\alpha \beta}$ (with $\alpha, \beta$ = A or B) such that $\epsilon_{\rm AA} = \epsilon_{\rm BB} \equiv  \epsilon$ and $\epsilon_{\rm AB} = \xi \epsilon$ with $\xi = 1.5$. $\epsilon$ and $\sigma$ represent in the following energy- and length-parameter. 

The system was simulated in (non-equilibrium) molecular dynamics (MD) simulations, using the LAMMPS package \cite{plimpton1995fast}. Working throughout in the canonical (NVT) ensemble a dissipative particle dynamics (DPD) thermostat was applied. Details about the DPD formalism can be found in  Ref. \cite{soddemann2003dissipative} and in Ref. \cite{Biswas2024} (see notably Eqs. (3) to (7) therein). The potential (\ref{gem4-equation}) was truncated at $r_{\rm c} = 2.2~\sigma$ where $\Phi(r_{\rm c}) = \epsilon \times 6.705 \times 10^{-11}$. In an effort to guarantee numerically reliable results we have considered (in most of our simulations) rather large ensembles, composed by $N = 65~536$ particles, assuming periodic boundary conditions in a cubic box (of volume $V$ and with side length $L = 32$). Throughout, the reduced, dimensionless number density, $\rho^\star = (N/V) \sigma^3$, was set to $\rho^\star = 2$. In an effort to study size effects of the results we have considered in our investigations ensembles of $N$ = 524~288 particles (corresponding to a value of $L = 64$ of a cubic box), $N$ = 65~536 (with $L = 32$), and $N$ = 8~192 (with $L = 16$).

Time $t$ and temperature $T$ are given in units of $t_0 = \sigma \sqrt{m/\epsilon}$ and $k_{\rm B} T/\epsilon$, respectively; we assume that both species have the same mass $m$, further $k_{\rm B}$ is the Boltzmann constant. The quantities $\epsilon$, $\sigma$, $m$, and $k_{\rm B}$ have been set to unity. 

For the initial configuration of the system the A and the B particles are distributed randomly in the simulation box. The system was first equilibrated over $10^6$ MD steps (with a step size of $\Delta t = 0.005 ~ t_0$) at the temperature $T=5.0$. Then the system is again simulated over $10^6$ $\Delta t$. Along this run, data was stored in intervals of $10^5$ time steps and the related particle positions were then stored as independent initial configurations for subsequent simulations at different temperatures $T$: for each of these $T$-values a simulation was launched from such a configuration, consisting of an equilibration phase (extending over $10^6$ $\Delta t$) and a subsequent production run (again over $10^6$ $\Delta t$). 

For the calculation of the mean square displacement, the velocity auto-correlation function, and the stress auto-correlation function -- for the related definitions see Section \ref{sec:results} -- we performed production runs in the microcanonical (NVE) ensemble: these simulations were started from equilibrated configuration produced in the NPT ensemble; subsequently the DPD thermostat was switched off. In the NVE ensemble the equilibrated systems were simulated over $2\times10^5$ $\Delta t$, data were stored every 100 $\Delta t$. 

We also exert Couette flow on the system via Lees-Edwards boundary conditions \cite{lees1972computer}. In our geometry the shear is applied along the $(x, z)$-plane in the $x$-direction. We have applied different shear rates $\dot \gamma$, ranging from $\dot \gamma = 10^{-5}$ to $\dot \gamma = 10^{-4}$ (namely, $10^{-5}$, $2\times10^{-5}$, $3\times10^{-5}$, $7\times10^{-5}$, $10^{-4}$), measured in units of $t_0$ (as defined above). When simulating the system under shear we started from equilibrium configurations obtained in previous runs and extended the simulation runs over $2~10^6 \Delta t$, exposing the system to shear forces; along these runs data are stored at every 10 $\Delta t$ for further analysis. Results (in terms of averages) are obtained once the system has reached its steady state by averaging over 300 different ensembles.

\section{Results}
\label{sec:results}

\subsection{Phase diagram and simulation snapshots}
\label{subsec:phase_diagram}

\begin{figure}[h!]
\centering
\includegraphics[width=0.45\textwidth]{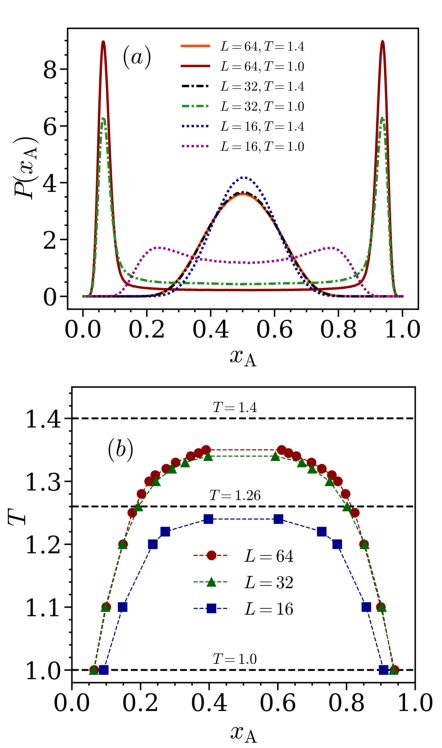}
\caption{Top panel (a): probability distribution, $P(x_{\text{A}})$ -- see text --  as a function of the concentration of A-particles, $x_\text{A}$, for selected temperatures $T$ (as labeled) and ensemble sizes (as labeled and specified via the edge length $L$ of the cubic simulation cell). Bottom panel (b): emerging phase diagram -- temperature $T$ vs. concentration $x_{\text{A}}$ -- for different ensemble sizes (as labeled in terms of $L$); the horizontal broken lines indicate those temperatures, for which snapshots are shown in Fig. \ref{fig:snapshots_fse}.}
\label{fig:pd-coex}
\end{figure}

Since we want to study the dynamic properties of our system in the vicinity of the critical point, we first have to trace out the phase diagram in a reliable manner. To this end we have used essentially the same approach that we have already applied in Ref. \cite{Biswas2024} and refer the reader for details to Section III.B of this contribution. In Fig. \ref{fig:pd-coex} 
we display in the top panel the probability distribution $P(x_{\text{A}})$, i.e., the distribution of the local concentration of particles A, for different temperatures and different sizes of the ensemble (as labeled). The positions of the peaks of $P(x_{\text{A}})$ define for a given temperature $T$ the location of the coexistence points at $T$ of the demixing transition; the emerging phase diagrams are shown in the bottom panel of Fig. \ref{fig:pd-coex} for different sizes of the ensemble (as labeled). Using the same method as presented in detail in Section III.B of Ref. \cite{Biswas2024}, we can estimate the critical temperature for different sizes of the ensemble; we obtain: $T_{\text{c}} = 1.243$ ($L$ = 16), $T_{\text{c}} = 1.342$ ($L$ = 32), and $T_{\text{c}} = 1.351$ ($L$ = 64). Since the values for the critical temperature differ only marginally for $L = 64$ and $L = 32$ we have opted to consider henceforward only the medium sized ensemble (i.e., for $L = 32$ and $N = 65~536$), which represents an excellent compromise between system size, accuracy, and computational effort. As will also be shown in Subsection \ref{subsec:structure} we will pursue an alternative path to a quantitative determination of the critical temperature which will lead to a consistent result.

\begin{figure}[h!]
\centering
\includegraphics[width=0.45\textwidth]{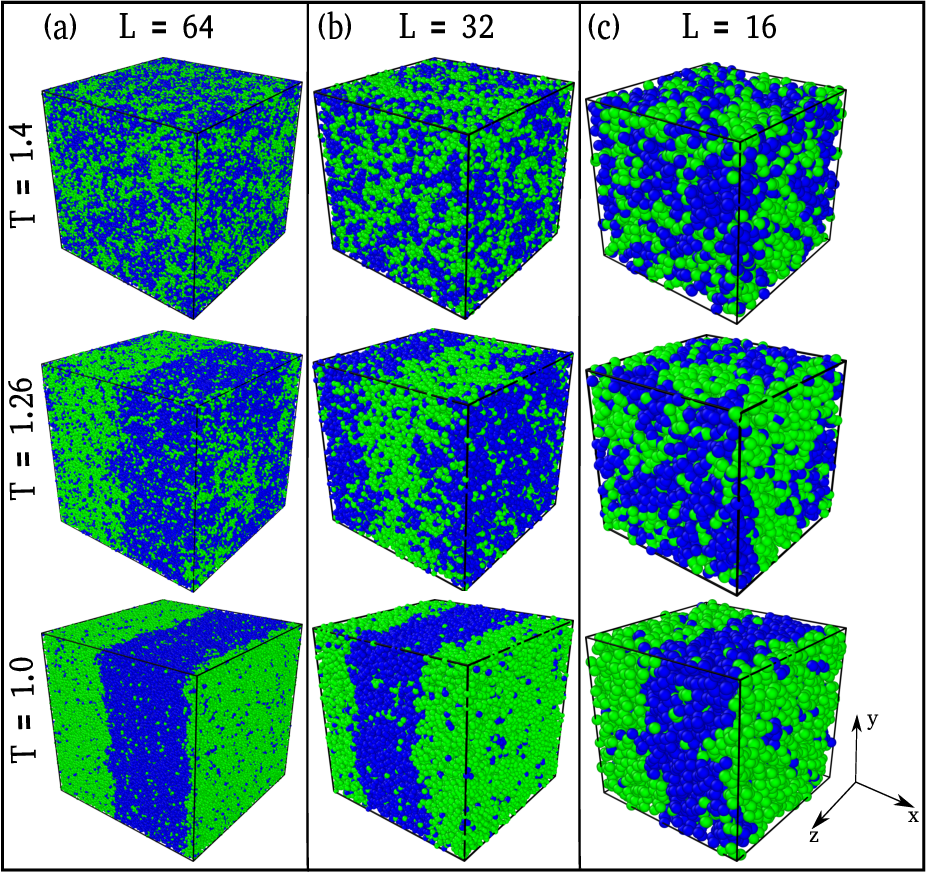}
\caption{Snapshots of the system in a cubic box (of edge length $L$) at the supercritical temperature $T = 1.4$ (top panels) and at the subcritical temperature $T = 1.0$ (bottom panels) for different ensemble sizes (as labeled in terms of $L$). The snapshots shown in the central panels are taken at $T=1.26$, i.e., a temperature which is subcritical for $L = 64$ and $L = 32$ and supercritical for $L = 16$. A- and B-particles are shown in blue and green, respectively. The related isotherms are shown in Fig. \ref{fig:pd-coex} as broken lines.}
\label{fig:snapshots_fse}
\end{figure}

Related snapshots of the system are shown in Fig. \ref{fig:snapshots_fse}, focusing on three ensemble sizes, namely $L = 64$, $L = 32$, and $L = 16$: these snapshots are displayed in the top row for a temperature $T = 1.4$ (which is supercritical for all system sizes) and in the bottom panels for a temperature $T = 1.0$ (which is subcritical for all system sizes). The central panels show snapshots at a temperature $T = 1.26$, which is subcritical for $L = 64$ and $L= 32$ and supercritical for $L = 16$ (see the related horizontal lines in panel (b) of Fig. \ref{fig:pd-coex}(b)). Throughout A and B particles are shown in different colours. At $T=1.4$ all systems are homogeneously mixed (albeit to a lesser extent for the smallest system), while at $T=1.0$  a clear phase separation is visible for all ensemble sizes considered. At the intermediate temperature of $T = 1.26$ a clear phase separation is visible for $L=64$ and $L=32$ (as the temperature is subcritical in both cases), while for $L=16$ (where the temperature lies slightly above the critical temperature) a clear onset towards phase separation is visible.

\subsection{Structure of the system}
\label{subsec:structure}

The structure of the system has been analysed in terms of the particle based partial radial distribution functions (RDFs) and different types of structure factors, tailored for binary systems.

The RDFs, $g_{\alpha \beta}(r)$ (with $\alpha, \beta$ = A or B), are defined as functions of the distance $r$ via \cite{hansenmcdonald}: 

\begin{equation}
    g_{\alpha \beta}(r) = \frac{N}{\rho N_\alpha N_\beta} \left<\sum_{i=1}^{N_\alpha} \sum_{j=1}^{N_\beta}{}^{'} \delta(r-|{\bf r}_i - {\bf r}_j|) \right> ;
    \label{partial_gr}
\end{equation}
the brackets denote an ensemble average. The prime indicates that for $\alpha = \beta$ the related contribution to the sum is suppressed in the case $i = j$. Further the partial structure factors are defined as functions of the wave vector ${\bf q}$ (with $\mid {\bf q} \mid = q)$ as \cite{hansenmcdonald, Bhatia1970structural}

\begin{equation} 
S_{\alpha \beta}(q) = \frac{(\delta_{\alpha \beta} + 1)}{2} \frac{1}{N} \sum_{k=1}^{N_\alpha} \sum_{l=1}^{N_\beta} \left<\exp[i {\bf q}  ({\bf r}_k - {\bf r}_l]] \right> .
\label{partial_sq}
\end{equation}
With these functions in mind one can define the concentration-concentration and the number-number partial structure factors, $S_{\rm cc}(q)$ and $S_{\rm nn}(q)$, respectively, via  \cite{hansenmcdonald, Bhatia1970structural};

\begin{eqnarray}
    S_{\rm cc}(q) &=& x_{\rm B}^2 S_{\rm AA}(q) + x_{\rm A}^2 S_{\rm BB}(q) - 2x_{\rm A}^2x_{\rm B}^2 S_{\rm AB}(q) \label{scc} \\
    S_{\rm nn}(q) &=& S_{\rm AA}(q) + 2 S_{\rm AB}(q) + S_{\rm BB}(q) .
\label{snn}    
\end{eqnarray}

\begin{figure}[h!]
\centering
\includegraphics[width=0.45\textwidth]{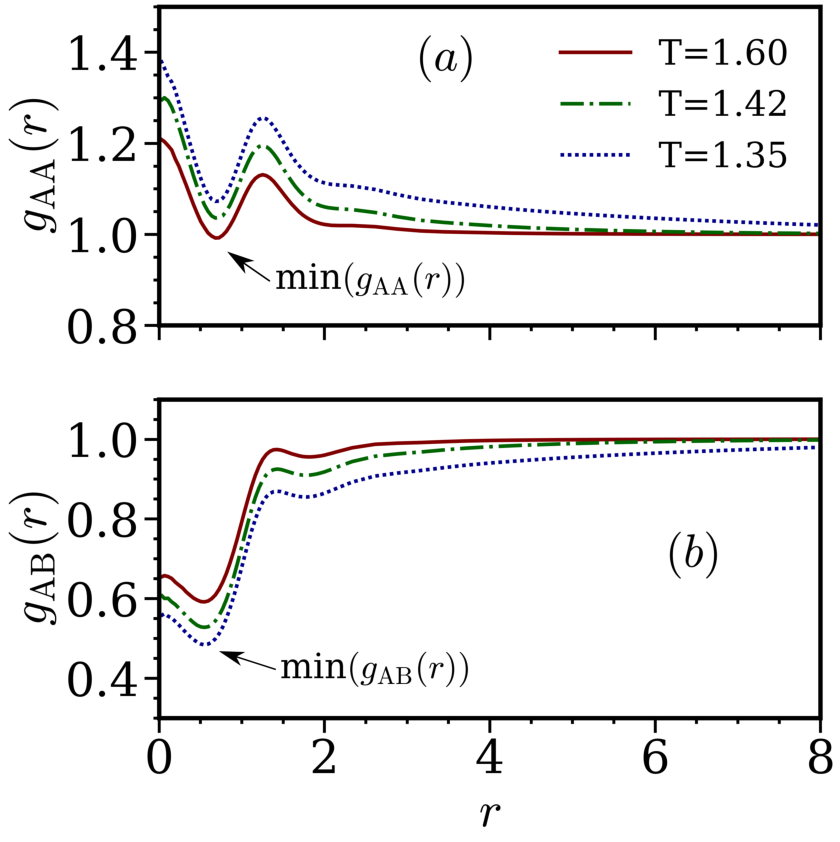}
\caption{Particle based partial RDFs $g_{\alpha \beta}(r)$ as defined in the text as functions of the distance $r$ for different (supercritical) temperatures (as labeled). The arrow points at the distance ${\rm min}(g_{\alpha \beta}(r))$ which is defined and discussed in the text. $\alpha \beta$ = AA -- panel (a), $\alpha \beta$ = AB -- panel (b). Throughout data were obtained for systems with $L = 32$.}
\label{fig:gr}
\end{figure}

In Fig. \ref{fig:gr} we show the partial RDFs $g_{\rm AA}(r) \equiv g_{\rm BB}(r)$ and $g_{\rm AB}(r)$  for different supercritical temperatures as labeled. The like RDFs $g_{\alpha \alpha}(r)$ show the expected behaviour: at short distances the functions are characterized by a sizable value at the origin, indicating cluster formation within the system, with the peak being more pronounced for temperatures in the vicinity of the critical temperature due to the increasing concentration fluctuations. At large distances the RDFs decay versus unity, the decay being the slower the more the temperature approaches the critical temperature. For the unlike RDF $g_{\rm AB}(r)$  we observe at short distances a value that lies distinctively below unity, reflecting the mutual repulsion between unlike particles; the impact of this repulsion is the stronger the more the system approaches the critical point. 

Since our ultrasoft particles tend to form clusters of overlapping particles we have also examined the structure of the system on the level of the emerging clusters. To this end two particles are considered to belong to the same cluster if their interparticle distance is less than the distance that corresponds to the first minimum in the $g_{\alpha \beta}(r)$; this distance -- termed ${\rm min}(g_{\alpha \beta}(r))$ -- is indicated by arrows in the panels of Fig. \ref{fig:gr}. In our investigations we have chosen ${\rm min}(g_{\alpha \beta}(r)) \simeq 0.63$. In addition we introduce the purity parameter $\pi$ which provides information about the purity of a cluster in terms of its occupancy by the two species of particles: for a given cluster $\pi$ is defined as the fraction of the dominant species (irrespective of A and B particles) with $\pi$ obviously ranging in the interval [0.5, 1].

\begin{figure}[h!]
\centering
\includegraphics[width=0.45\textwidth]{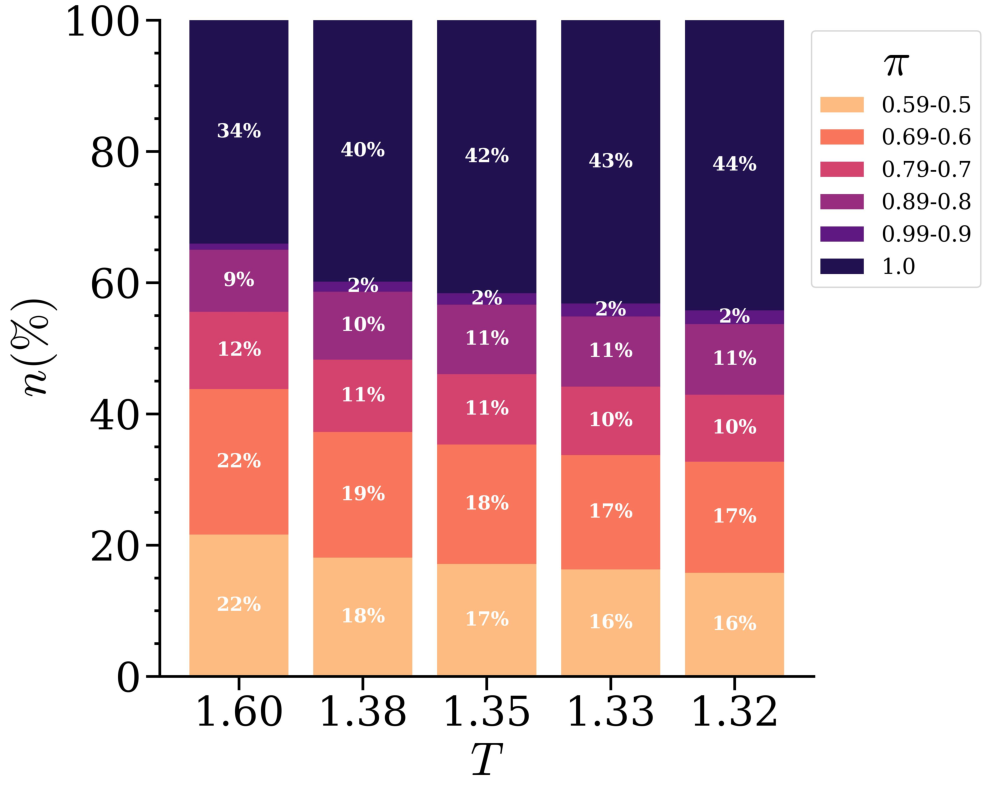}
\caption{Stacked bar chart of the percentage $n$ of clusters, characterized by a specific value of the purity parameter $\pi$, which is defined in the text. The colour code on the right hand side of the figure specifies the different ranges of $\pi$ that have been considered, ranging from the pure case (i.e., $\pi = 1$) to the perfectly mixed case (i.e., $\pi = 0.5$). Several super- and subcritical temperatures have been considered (as labeled). Isolated particles have not been considered for this analysis. Throughout data were obtained for $L = 32$.}
\label{fig:cluster_purity}
\end{figure}

A quantitative analysis of $\pi$ is shown in Fig. \ref{fig:cluster_purity}: in a stacked bar chart we display $n$, i.e., the percentage of clusters with a given $\pi$-value; to this end the interval $\pi \in [0.5, 1]$ is divided into different sub-intervals as specified via the colour code of Fig. \ref{fig:cluster_purity}. Different temperatures have been considered for this analysis. We observe that with decreasing temperature -- from the supercritical $T = 1.60$ down to the subcritical $T = 1.32$ -- a distinct increase in the percentage of (essentially) pure clusters (i.e., $\pi \simeq 1$) is observed: the fraction of pure clusters increases steadily from $34\%$ (at $T=1.60$) to $44\%$ (at $T=1.32$). For $\pi < 1$ (i.e., for more or less mixed clusters) five intervals have been introduced -- see labels in Fig. \ref{fig:cluster_purity}. 
In contrast to the pure clusters we observe for the mixed clusteers (i.e., for $\pi \lesssim 0.9$) for each $\pi$-window a decrease of $n$ of the related clusters with decreasing temperature. For strongly mixed clusters (i.e., for $0.5  \le \pi \le 0.59$) the decrease of $n$ is very pronounced: it drops from 22 \% (at $T = 1.6$) to 16 \% (at $T = 0.32$). All the above said quantifies and fully confirms the expected enhanced tendency towards demixing in the binary mixture as the temperature is decreased from super- to subcritical temperatures. 

With the clusters being readily defined we can now introduce the cluster based RDFs, specified by a superscript 'c' and defined via 

\begin{equation}
g^{\rm c}_{\alpha \beta}(r) = \left< \frac{V}{n_\alpha n_\beta} \sum_{i=1}^{n_\alpha} \sum_{j=1}^{n_\beta}{}^{'} \delta(r-|{\bf r}^{\rm CM}_i - {\bf r}^{\rm CM}_j|) \right> ;
\label{partial_gr_cm}
\end{equation}
these functions correlate the centers of mass ${\bf r}^{\rm CM}$ of the clusters, defined as

\begin{equation}
    {\bf r}^{\rm CM} = \frac{1}{N_{\rm c}} \sum_{i=1}^{N_{\rm c}} {\bf r}_i ;
\end{equation}
here the sum is extended over the $N_{\rm c}$ particles that pertain to a given cluster. The $n_\alpha$ and $n_\beta$ are the numbers of $\alpha$- or $\beta$-type clusters in the system (see below). The brackets in Eq. (\ref{partial_gr_cm}) represent an ensemble average and the prime in the second sum indicates that the contribution to the sum is suppressed if $i = j$. 
\begin{figure}[h!]
\centering
\includegraphics[width=0.36\textwidth]{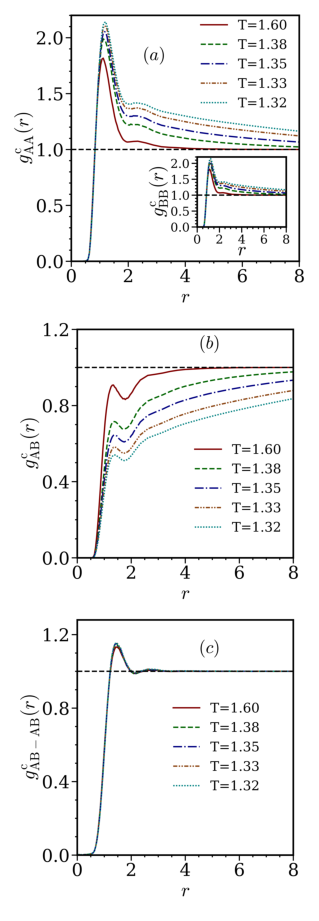}
\caption{Cluster based partial RDFs $g^{\rm c}_{\alpha \beta}(r)$ as defined in the text as functions of the distance $r$ between the COM of two clusters for different super- and subcritical temperatures (as labeled). $\alpha \beta$ = AA -- panel (a), $\alpha \beta$ = AB -- panel (b), $\alpha \beta$ = AB-AB -- panel (c). Isolated particles have not been considered for this analysis. Throughout data were obtained for $L = 32$.}
\label{fig:gr_cluster}
\end{figure}

The index combination 'AA' (or 'BB') denotes the correlations between pure A (or pure B) clusters, i.e., $\pi = 1$. 'AB' specifies correlations between pure A and pure B clusters, while 'AB-AB' specify correlations between mixed clusters (i.e., for $\pi < 1$). Isolated particles have not been included in this analysis (see also remark further down). The respective RDFs are depicted in the three panels of Fig. \ref{fig:gr_cluster}. All RDFs are shown for a temperature range covering both super- and subcritical temperatures.  It should be noted that all cluster based RDFs vanish for small $r$-values indicating that the cluster are -- irrespective of their individual composition -- characterized by an effective, strongly repulsive interaction. They can thus be viewed as mutually repulsive effective particles. 

Starting at $T = 1.6$ we observe a significant increase in the main peak of $g^{\rm c}_{\rm AA}(r) \equiv  g^{\rm c}_{\rm BB}(r)$ as the temperature is decreased, providing evidence of a strong increase in the spatial correlations between pure, like clusters. Concomitantly the decay of this function towards unity at large distances is the slower the lower the temperature: thus at lower temperatures the correlations between pure, like clusters pertains over even longer distances as compared to higher temperatures. 

In contrast, $g_{\rm AB}^{\rm c}(r)$ assumes for all temperatures investigated values that are smaller than unity; the hight of the local peak $g^{\rm c}_{\rm AB}(r \simeq 1.32) \simeq 0.91$ at $T = 1.6$ drops down to $g^{\rm c}_{\rm AB}(r \simeq 1.32) \simeq 0.54$ at $T = 1.32$, indicating that the correlations between pure, unlike clusters drops drastically as we pass the critical temperature. Further, the fact that $g^{\rm c}_{\rm AB}(r)$ tends considerably slower towards unity as the temperature is decreased provides evidence of an increasingly reduced correlation of pure, unlike clusters at larger separations.  

Surprisingly the cluster based RDFs between mixed clusters, $g^{\rm c}_{\rm AB-AB}(r)$, does not show any temperature dependence, indicating that these particular correlations are stable over a surprisingly broad temperature. This observation obviously also includes the near-critical region where this RDF is obviously bare of any sign of criticality. 

It should be noted that including in the above analysis also isolated particles (which can be viewed as singly occupied, pure A or B clusters) leads to small quantitative changes in the RDFs but the qualitative conclusions remain unaffected.  Thus we refrain from a more detailed presentation of these results.

\begin{figure}[h!]
\centering
\includegraphics[width=0.45\textwidth]{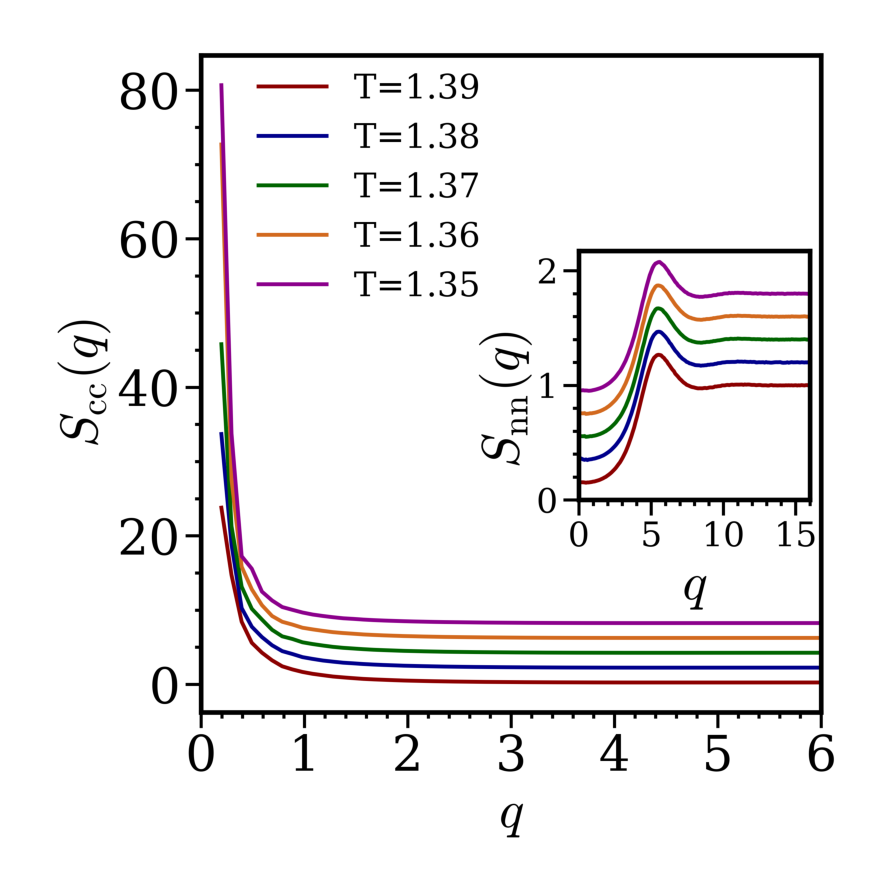}
\caption{Concentration-concentration structure factors $S_{\rm cc}(q)$ of the system as functions of the wave-vector $q$ for different temperatures (as labeled). For better visibility curves are shifted by two units. The inset shows the number-number structure factors $S_{\rm nn}(q)$ of the system as functions of the wave-vector $q$ for different temperatures (as labeled). Here, curves are shifted by 0.2 units for better visibility. Throughout data were obtained for systems with $L = 32$.}
\label{fig:snn-scc}
\end{figure}

\begin{figure}[h!]
\centering
\includegraphics[width=0.45\textwidth]{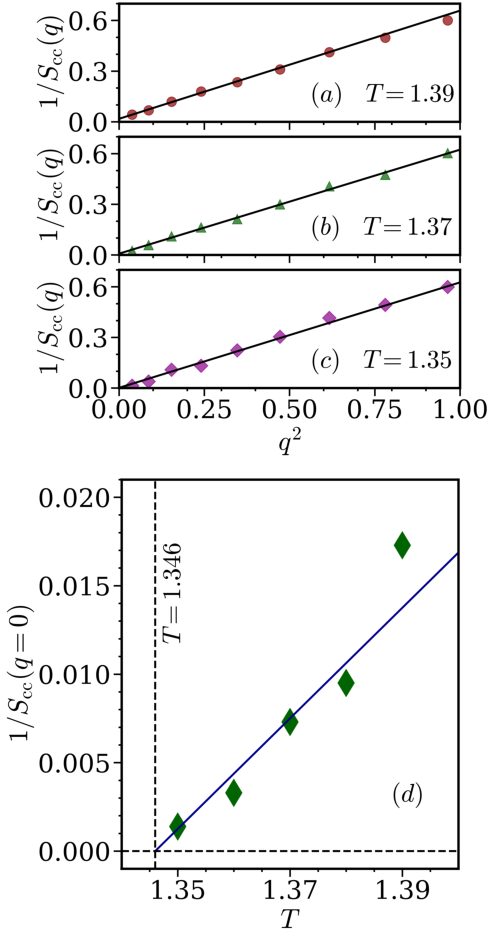}
\caption{Panels (a) to (c): Ornstein-Zernike plot \cite{oz1914,oz1918,fisher1964correlation} of the concentration-concentration structure factor $S_{\rm cc}(q)$, i.e., $1/S_{\rm cc}(q)$ vs. $q^2$ in the low-$q$ regime for different temperatures (as labeled). The black lines indicate the linear fit of to the data for small values of $q$. Bottom panel: $1/S_{\rm cc}(q=0)$ -- as extrapolated from the data in the limit $q \to 0$ --  versus temperature $T$; the blue line is a linear fit to these data. The dashed, vertical black line indicates the $T$-value where the blue line has its zero, corresponding to the critical temperature $T_{\rm c}$ -- see text. Throughout data were obtained for systems with $L = 32$.}
\label{fig:sccq0}
\end{figure}

$S_{\rm cc}(q)$ and $S_{\rm nn}(q)$ -- as defined in Eqs. (\ref{partial_sq}) to (\ref{snn}) -- are shown in Figs. \ref{fig:snn-scc} and \ref{fig:sccq0}. In the former case we see the expected divergence of $S_{\rm cc}(q)$ in the vicinity of the critical point as $q$ tends towards zero which emerge as a consequence of concentration fluctuations near criticality; for better visibility the curves of $S_{\rm cc}(q)$ are shifted vertically by two units. In contrast, $S_{\rm nn}(q)$ (see inset of Fig. \ref{fig:snn-scc}) does not show any hallmark in the vicinity of the critical point. In an effort to elucidate on a more quantitative level the divergence of $S_{\rm cc}(q)$ we show $1/S_{\rm cc}(q)$ as a function of $q^2$ in panels (a) to (c) of Fig. \ref{fig:sccq0} for small values of $q$, also known in literature as the Ornstein-Zernike plot; in this region $S_{\rm cc}(q)$ should behave as \cite{oz1914,oz1918,fisher1964correlation}

\begin{equation}
    S_{\rm cc} (q) \simeq k_{\rm B} T \frac{\chi_{\rm T}}{1+\xi^2 q^2} ,
\label{OZ}    
\end{equation}
where $\chi_{\rm T}$ and $\xi$ are the isothermal susceptibility and static correlation length, respectively. From these panels we observe for temperatures slightly above the critical point the predicted linear behaviour of $1/S_{\rm cc}(1)$ vs. $q^2$ within high accuracy -- see Eq. (\ref{OZ}). 

As can be seen in panels (a) to (c) of Fig. \ref{fig:sccq0} we can obtain via linear interpolation of the data the value $\lim_{q \to 0}[1/S_{\rm cc}(q)]$ with high accuracy; the related data are accumulated in Table \ref{tab:sccq=0} and are presented in panel (d) of Fig. \ref{fig:sccq0}. This analysis offers an alternative method to estimate the critical temperature $T_{\rm c}$ as follows: as predicted by theory \cite{Bhatia1970structural}, the divergence of $S_{\rm cc}(q=0)$ -- or, alternatively, the vanishing of $[1/S_{\rm cc}(q=0)]$ -- should occur at the critical temperature, which we can now extract by extrapolating (via a linear interpolation function) the values of $[1/S_{\rm cc}(q=0)]$ as a function of temperature towards zero, displayed by the vertical broken line in panel (d) of Fig. \ref{fig:sccq0}. According to this extrapolation the critical temperature occurs at $T_{\rm c} \simeq 1.346$, a value which agrees nicely with $T_{\rm c} = 1.342$ as obtained from the analysis of the phase diagram (see Subsection \ref{subsec:phase_diagram}).

\begin{table}[h!]
    \centering
    \begin{tabular}{|P{4cm}|P{4cm}|} \hline 
         \rowcolor{Gray}
         $T$ & $1/S_{\rm cc}(q=0)$\\ \hline
         $1.35$ & $0.00138$ \\ \hline
         $1.36$ & $0.00329$ \\ \hline
         $1.37$ & $0.00729$ \\ \hline
         $1.38$ & $0.00949$ \\ \hline
         $1.39$ & $0.01728$ \\ \hline
    \end{tabular}
    \caption{Values of $\lim_{q \to 0} [1/S_{\rm cc}(q)]$, obtained as extrapolated values from the data shown in panels (a) to (c) of Fig. \ref{fig:snn-scc} for different values of temperature $T$.}
    \label{tab:sccq=0}
\end{table}

\subsection{Diffusion}
\label{subsec:diffusion}

In an effort to calculate the diffusion properties of the particles within the system we start with the mean-square displacement (MSD). Since the mixture is equimolar and size symmetric both species have the same diffusive behaviour; thus we suppress for simplicity in the following the species index. The MSD is defined as

\begin{eqnarray}
    (\Delta r(t))^2 = \frac{1}{N} \sum_{i=1}^{N} \left<|
    {\bf r}_{i}(t) - {\bf r}_{i}(0)|^2\right> ,
    \label{msd}
\end{eqnarray}
where ${\bf r}_{i}(t)$ is the position of particle $i$ at time $t$. The brackets indicate an ensemble average, realized in practice via gliding windows along the simulation run. 

The MSD itself shows for all temperatures investigated -- be it super- or subcritical -- the expected ballistic and diffusive behaviour for small and large $t$-values, respectively. These features are not only confirmed via linear fits in the respective regimes but also via an explicit calculation of the instantaneous slope $\alpha$ of the MSD, defined as

\begin{eqnarray}
    \alpha = \frac{d}{d(\ln t)} \ln (\Delta r(t))^2 .
    \label{instantslope}
\end{eqnarray}
Albeit not displayed in this contribution, $\alpha$ assumes within high accuracy the values of unity and two within the two above mentioned time-windows. 

\begin{figure}[h!]
\centering
\includegraphics[width=0.45\textwidth]{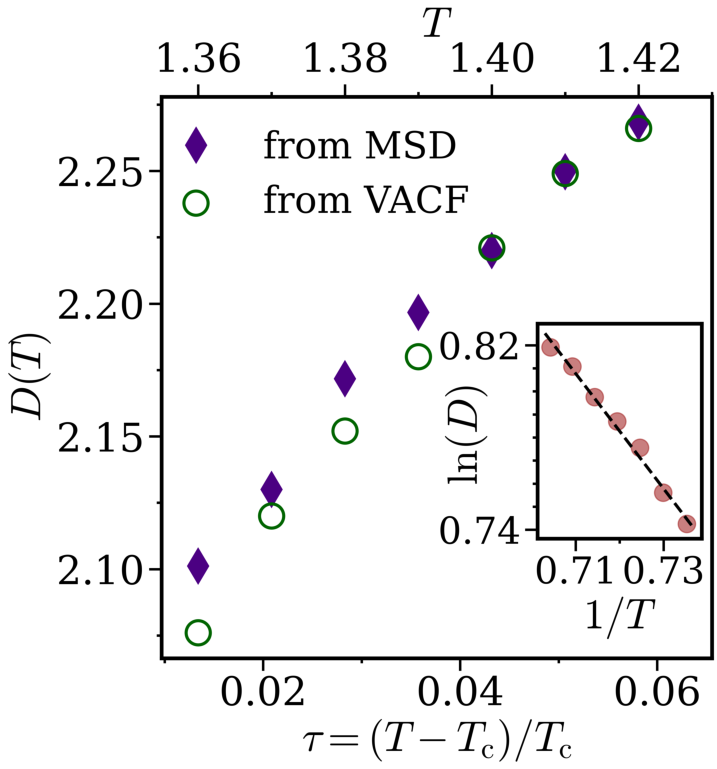}
\caption{Diffusion constant $D = D(T)$ as a function of temperature $T$ (horizontal top scale) and of $\tau =(T - T_{\rm c})/T_{\rm c}$, i.e., the relative deviation of the temperature from the critical temperature (horizontal bottom scale). Data with different symbols have been obtained via the MSD or via integrating the VACF (as labeled) -- see text. Inset: $\ln D$ as a function of $1/T$ (symbols) along with a linear fit (broken line) which indicates an Arrhenius-law behaviour (see text). Throughout data were obtained for systems with $L = 32$.} \label{fig:diffusion}
\end{figure}

Also for the diffusion constant, $D = D(T)$ we can suppress -- due to symmetry of the model -- the species index; this quantity has been extracted from the MSD via the well-known Stokes-Einstein relation \cite{allen2017computer}

\begin{eqnarray}
D = \frac{1}{6t} \lim_{t\to\infty} (\Delta r(t))^2  .
\label{diffconst}
\end{eqnarray}
$D(T)$ is displayed for different temperatures in Fig. \ref{fig:diffusion}, both as a function of temperature $T$ (horizontal top scale) as well as of the relative temperature deviation from the critical temperature, defined via $\tau = (T - T_{\rm c})/T_{\rm c}$ (horizontal bottom scale).

The inset of Fig. \ref{fig:diffusion} shows $\ln D(T)$ as a function of $1/T$. The excellent fit of this data by a straight (broken) line indicates that the diffusion constant follows with high accuracy the Arrhenius law \cite{Arrhenius1889, alberty1997physical}, i.e., 

\begin{eqnarray}
D = D_0 \exp{[-E_{\rm a}/(k_{\rm B}T)]} .
\label{arrlaw}
\end{eqnarray}
In the above relation $E_{\rm a}$ is the activation energy. The actual value as obtained from the fitting procedure is $E_{\rm a} = 2.492$. In an effort to compare this value to the related result of a binary, size-symmetric, equimolar LJ mixture (at a reduced density $\rho = 1$) we have carried out related simulations for this system. From these runs we can extract a value $E^{\rm LJ}_{\rm a} = 2.731$ (in units of the LJ $\epsilon$), i.e., a value that is by 10 \% higher than the GEM-value. This indicates that the GEM mixture at hand has throughout a higher diffusivity, i.e., that the GEM particles are -- as a consequence of their softness -- considerably more diffusive than the LJ particles. 

Alternatively one can calculate $D = D(T)$ by integrating the velocity auto-correlation function (VACF), $\Psi(t)$, i.e.,  

\begin{eqnarray}
D = \frac{1}{3} \int_{0}^{\infty} \Psi(t) d t
\label{diffvacf}
\end{eqnarray}
where $\Psi(t)$ is given by 

\begin{eqnarray}
\Psi(t) = \frac{\sum_{i=1}^N \left < {\bf v}_i(t) \cdot {\bf v}_i(0)\right >}
{\sum_{i=1}^N\left< {\bf v}_i(0) \cdot {\bf v}_i (0)\right >} .
    \label{vacf}
\end{eqnarray}
In practice the integral in Eq. (\ref{diffvacf}) has been extended over $10^4$ time steps, similar as for Eq. (\ref{msd}) the brackets indicate ensemble averages, realized in practice via gliding windows along the simulation run. The related values obtained for $D(T)$ for different temperatures have been added to the data of $D$ (obtained via the MSD); they are all accumulated in Fig. \ref{fig:diffusion}. The two sets of data show a very good agreement. 

\subsection{Viscosity}
\label{subsec:viscosity}

We eventually proceed to the last transport coefficient investigated in this study, the shear viscosity $\eta$. To this end we employ the Green-Kubo formalism \cite{Green1954, Kubo1957}, with the related equations based on the stress auto-correlation function (SACF). In a later stage (see Subsec. \ref{subsec:shear}) we will expose the system to external shear (employing a constant shear rate) which offers the possibility to determine the shear viscosity in an alternative manner. 

The formalism starts from the SACF $C_{\rm s}(t)$ which is calculated from the data available from the computer simulations via \cite{allen2017computer}

\begin{equation}
    C_{\rm s}(t) = \left< \sigma_{xz}(t) \sigma_{xz}(0) \right >
    \label{sacf}
\end{equation}
where $\sigma_{xz}(t)$ is the off-diagonal element of the stress tensor at time $t$, i.e., 

\begin{equation}
\begin{split}
    \sigma_{xz}(t) &= \frac{1}{V} \sum_{i=1}^{N}\Big[m v_{ix}(t) v_{iz}(t) \\
    &\quad + \sum_{i<j}^{N} |x_i(t) - x_j(t)| F_z(|{\bf r}_i(t) - {\bf r}_j(t)|) \Big]
\end{split}
\label{stress}
\end{equation}
here $v_{ix}(t)$ and $v_{iz}(t)$ are the $x$- and the $z$-components of the velocity of particle $i$; further,  $F_{z}(r)$ is $z$-component of the force acting between particles $i$ and $j$ (located at positions ${\bf r}_i(t)$ and ${\bf r}_j(t)$). The first term in Eq. (\ref{stress}) represents the kinetic contribution to the stress tensor, while the second term is the virial contribution to $\sigma_{xz}(t)$. It should be noted that for a LJ system the kinetic part to the stress tensor is negligible, while in the case of GEM particles both contributions are of comparable size.

The SACF (not shown here) smoothly decays as a function of time $t$ to zero at all temperatures investigated. 

\begin{figure}[h!]
\centering
\includegraphics[width=0.45\textwidth]{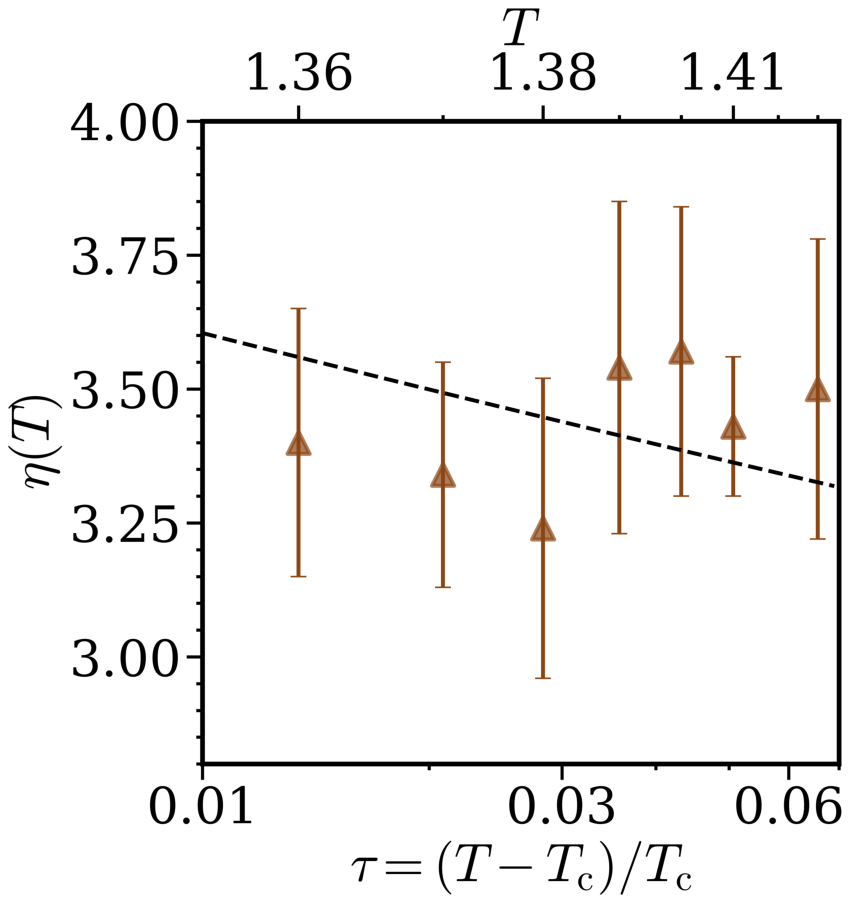}
\caption{Symbols (with error bars) -- viscosity $\eta = \eta(T)$, calculated within the Green-Kubo formalism via integrating the SACF over time (see text) as a function of temperature $T$. Dashed line -- predicted theoretical curve \cite{mistura1975, luettmer1995, luettmer1996} to which the simulation data have been fitted -- see Eq. (\ref{fitvisc}) and text.}
\label{fig:sacf-gkviscosity}
\end{figure}

From the SACF the viscosity $\eta$ can be determined via the Green-Kubo relation \cite{allen2017computer}:

\begin{eqnarray}
\eta = \eta(T) = \frac{V}{k_{\rm B} T} \int_{0}^{\infty} \left < \sigma_{xz}(t)\sigma_{xz}(0) \right> dt .
\label{gkvisc}
\end{eqnarray}
In practice the integral in Eq. (\ref{gkvisc}) has been extended over $10^4$ time steps. The related values of $\eta$ are displayed as a function of temperature $T$ in Fig. \ref{fig:sacf-gkviscosity} (both for absolute temperature values as well as for deviation of the temperature from the critical temperature, $\tau$ -- see the two horizontal scales). Note that -- in view of the above said and in striking contrast to the LJ system -- the values obtained for $\eta$ differ strongly if the kinetic contribution to the stress tensor -- see Eq. (\ref{stress} -- is included or not.

The temperature dependence of $\eta(T)$ can be fitted in the vicinity of the critical point via the following functional form: \cite{mistura1975, Sengers1985, luettmer1995, luettmer1996}, with $\tau = (T - T_{\rm c})/T_{\rm c}$

\begin{eqnarray}
    \eta \simeq \eta_{0}\tau^{-\nu x_{n}} .
     \label{fitvisc}
\end{eqnarray}
This relation is based on the following concepts: (i) in the critical region the viscosity follows the law  $\eta \simeq \eta_0 \xi^{x_n}$ \cite{ferrell1985, Sengers1985,  luettmer1995, luettmer1996, bhattacharjee1998,hao2005} with $\eta_0$ being the amplitude and $\xi$ being the static correlation length; further $x_n$ is the critical exponent of the viscosity. Since $\xi$ diverges near criticality via $\xi \simeq \tau^{-v}$ \cite{stanleybook1971, PathriaBeale2021, fisher1974renormalization} with $\nu$ being the well-known critical exponent of the correlation length, one arrives at the above relation. Within the Ising 3D universality class $\nu \simeq 0.629$ \cite{hasenbusch1999critical, Butera2002critical, pelissetto2002critical, subir2006static,  hasenbusch2010finite, kazmin2022critical}. Due to its smallness, the actual value of $x_n$ is rather difficult to identify and the values reported in literature are quite diverse. For this contribution we have used $x_n = 0.068$ \cite{ferrell1985,bhattacharjee1998, hao2005}. 

Fitting our data to the scaling law Eq. (\ref{fitvisc}) and using the above values of $\nu$ and $x_n$ in close vicinity of the critical point (i.e., for $\tau \in [0.01, 0.06]$) we obtain for the amplitude $\eta_0 \simeq 2.961$. This value is considerably smaller than the related LJ-value, $\eta_0^{\rm LJ} \simeq 3.87$ \cite{subir2006static}, indicating that the GEM system is for obvious reasons considerably less viscous than the related LJ mixture. Data for $\eta(T)$ are shown in Fig. \ref{fig:sacf-gkviscosity} along with the functional form specified in Eq. (\ref{fitvisc}), using the above mentioned numerical parameters; error bars are obtained by considering $\eta(T)$ values from different independent runs.

With the diffusion constant $D(T)$ and the viscosity $\eta(T)$ at hand we can calculate the Stokes-Einstein (SE) diameter, $d_{\rm SE} = d_{\rm SE}(T)$ via \cite{subir2003transport, subir2006static} (suppressing again the species index),

\begin{eqnarray}
    d_{\rm SE}(T) = \frac{T}{2 \pi \eta(T) D(T)} .
    \label{SE_dia}
\end{eqnarray}
The SE diameter represents in some manner the effective diameter of a (spherical) particle diffusing in a fluid of viscosity $\eta$.

\begin{figure}[h!]
\centering
\includegraphics[width=0.45\textwidth]{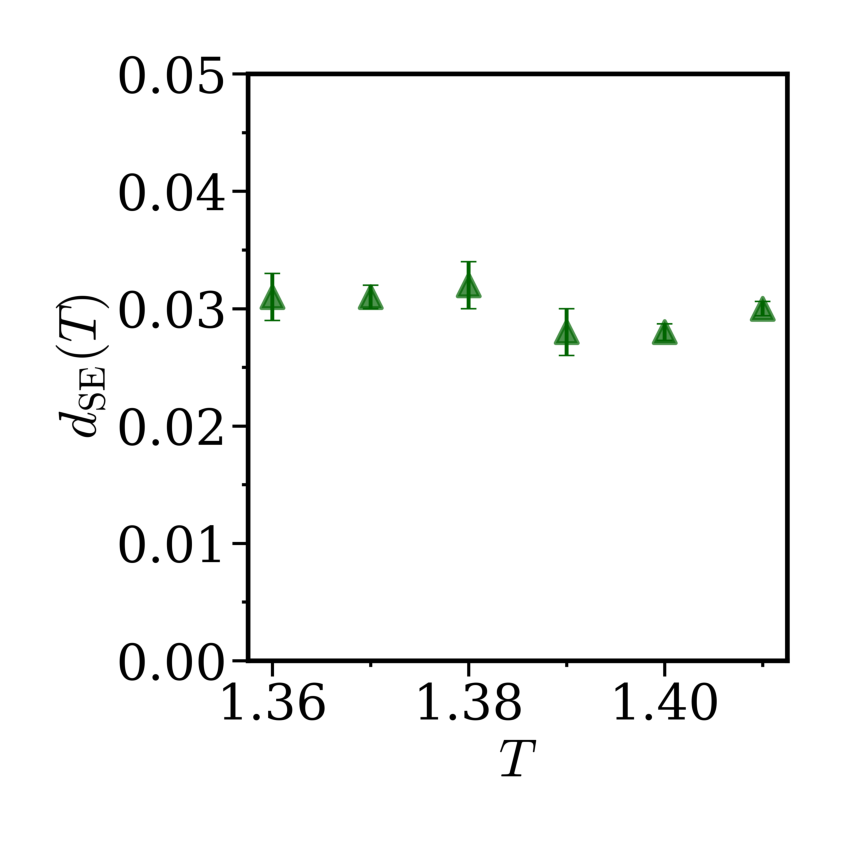}
\caption{Symbols (with error bars) -- Stokes-Einstein diameter $d_{\rm SE}(T)$ as calculated via Eq. (\ref{SE_dia}), as a function of temperature $T$.} 
\label{fig:stokes_einstein_diameter}
\end{figure}

$d_{\rm SE}$ is shown in Fig.~\ref{fig:stokes_einstein_diameter} as a function of temperature. We observe that this quantity remains essentially constant over the range of the considered temperatures. The actual value of $d_{\rm SE} \simeq 0.03$ is substantially smaller than the related value of the SE diameter of a LJ system ($d_{\rm SE}^{\rm LJ} \simeq 1$) \cite{subir2006static, subir2003transport}. This large discrepancy arises obviously from the ultrasoft nature of the GEM-4 particles. 

\subsection{The system under shear}
\label{subsec:shear}

In an effort to have an alternative access to the rheological properties of our system -- notably to the viscosity -- we have exposed our system to external shear forces (for details see Sec. \ref{sec:model_methods}). The values of the shear rates $\dot \gamma$ have been carefully chosen: $\dot \gamma$ assumes values that are sufficiently small, guaranteeing that the system remains within the linear response regime under the applied shear.

The shear stress $\sigma_{xz}(t)$ has been calculated via the Irving-Kirkwood expression, given in Eq. (\ref{stress}). According to the underlying formalism, the  viscosity is then obtained via

\begin{equation}
\eta = \frac{\left<\sigma_{xz}\right>}{\dot \gamma}
\label{eta_shear}
\end{equation}
where the brackets denote (i) an average over 300 independent runs and (ii) a time average once the system has reached a steady state; to be more specific the last 10$^6$ $\Delta t$ time steps of a simulation run (extending 2 $10^6$ $\Delta t$) have been considered for this time average, i.e., when the system has definitely reached a steady state. 
\begin{figure}[h!]
\centering
\includegraphics[width=0.45\textwidth]{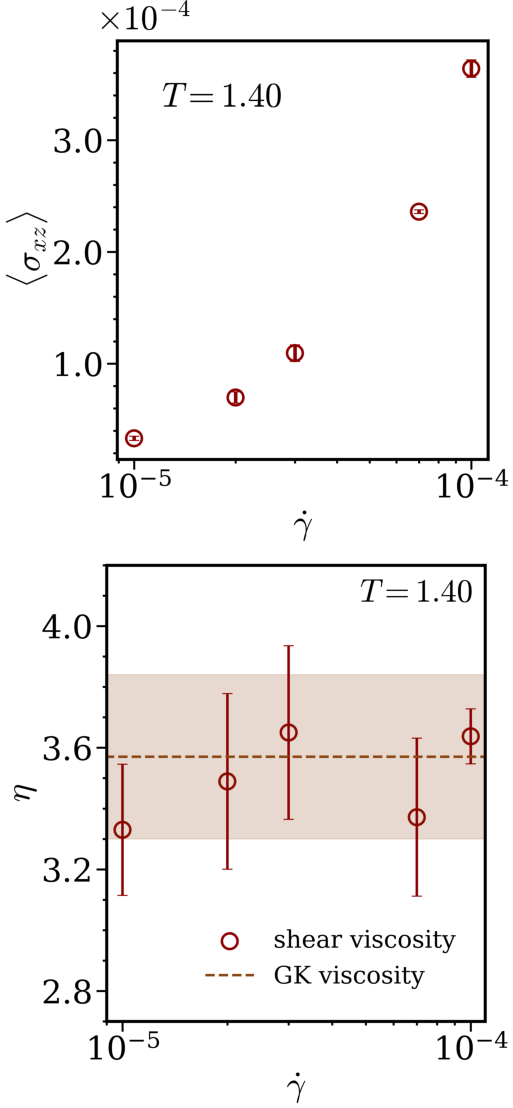}
\caption{Top panel: averaged stress, $\langle \sigma_{xz} \rangle$ (see text), as a function of the shear rate $\dot \gamma$ at $T=1.40$. Bottom panel: viscosity $\eta$ as a function of shear rate at $T=1.40$, as obtained from shear experiment via Eq. (\ref{eta_shear}) -- for details see text.  The horizontal dotted lines indicates the viscosity as obtained for $T = 1.40$ from the Green-Kubo (GK) formalism -- see Eq. (\ref{gkvisc}) -- the shaded regions covers the corresponding error bar -- see Fig. \ref{fig:sacf-gkviscosity}.}
\label{fig:shear-stress-viscosity-kv}
\end{figure}

Related data are compiled in Fig. \ref{fig:shear-stress-viscosity-kv}. The top panel shows $\left< \sigma_{xz} \right>$ for the selected $\dot \gamma$-values, calculated for $T = 1.40$, i.e., at a temperature that is in reasonable vicinity to the critical point. As expected the averaged stress shows a monotonically growing behaviour with increasing shear rate. The ensuing viscosity $\eta$ is shown in the bottom panel of this figure, calculated for the five $\dot \gamma$-values considered: $\eta$ varies only slightly with increasing shear rate as it can be expected within the linear response regime. Further the viscosity as determined from the shear experiments shows a good agreement with the related value obtained for $T = 1.40$ within the (static) Green-Kubo formalism -- see Eq. (\ref{gkvisc}): this agreement validates our approach for determining the viscosity via shear experiments.

\section{Summary and discussions}
\label{sec:summary}

In this contribution we have studied in detail a variety of properties of a binary, equimolar, size-symmetric mixture of ultrasoft particles as we lower the temperature from supercritical to subcritical temperatures; paraticle species are identified via the indices A and B. The particles interact via the ultrasoft ``generalized exponential model of index 4'' (GEM4): the like attractions are assumed to be identical, while the unlike interaction is a scaled version of the like interactions, introducing the scaling parameter $\xi = 1.5$. Investigations are based on extensive molecular dynamics (MD) simulations both in- and out-of-equilibrium (taking benefit of the LAMMPS package). After several investigations a system size of $N = 65~536$ particles confined in a cubic box with periodic boundary condition has been identified as a viable compromise between computational costs and numerical accuracy. Simulations have been performed both in the canonical (introducing a heat bath) as well as in the microcanonical ensemble. Further the system has been exposed to external shear forces (implemented via Lees-Edwards boundary conditions) introducing a shear rate $\dot \gamma$ of suitable strength.

After having localized the phase boundaries of the coexisting phases and the localisation of the critical point (in the latter case we have used two alternative, highly consistent routes) of the system we have investigated on a more qualitative level the cluster formation of the system, which -- for subcritical temperatures -- has two origins: the expected concentration fluctuations and the formation of clusters as a typical feature of GEM4 particles. The radial distribution functions (RDFs) of the individual particles show the expected behaviour for the correlations between like and the unlike particles as we lower the temperature to subcritical temperatures. Defining in a suitable manner cluster critria we have analysed in detail the emerging clusters with respect to their composition, introducing thereby a suitably defined purity parameter $\pi$. Based on different cluster based RDFs we find that the correlations between pure (A or B) clusters experience a pronounced increase in their spatial correlations as we pass the critical temperatures, while -- concomitantly -- the correlations between unlike, pure clusters are characterized by a serious decrease in their spatial correlations. 

We have furthermore calculated the diffusion constant $D = D(T)$ via two different, complementary routes (with excellent internal consistency) and have demonstrated that the diffusion constant shows -- not surprisingly -- for the GEM4 mixture shows an Arrhenius type of behaviour. The activation energy -- which represents in the Arrhenius law as a fitting parameter -- is found to be by 10 \% smaller than the value of a related, binary, equimolar, size-symmetric Lennard-Jones (LJ) mixture. Further we have calculated the viscosity $\eta =  \eta(T)$ via the stress auto-correlation function within the Green-Kubo formalism; this quantity shows in the vicinity of the critical point a typical power-law divergence. Using the related exponents (known within the Ising 3D universality class) we obtain within this framework an amplitude $\eta_0$ which is considerably smaller than the related LJ value, indicating that the GEM system is less viscous than a LJ mixture. Also the Stokes-Einstein diameter, which can be calculated directly from $D(T)$ and $\eta(T)$ is -- nearly by a factor 30 (!) -- smaller than the related value of the above mentioned LJ mixture, expressing thereby the softness of the GEM particles.

Our results provide -- along with data presented in a previous publication \cite{Biswas2024} -- a profound insight into the static and dynamic properties of a binary, equimolar, size-symmetric mixture of ultrasoft particles along with a detailed comparison to a related LJ system.

\section*{Acknowledgements}

TB is grateful to Emanuela Bianchi (TU Wien) for valuable  and helpful discussions. GK acknowledges financial support of the Austrian Science Foundation (FWF) under Proj. No. PIN8759524. The computational results have been achieved using the Austrian Scientific Computing (ASC) infrastructure under Project No. 71263. The authors acknowledge TU Wien Bibliothek for financial support through its Open Access Funding Programme.

\newpage








\bibliography{transport}

\end{document}